\documentclass[aps,english,prb,twocolumn,showpacs,superscriptaddress]{revtex4}
\usepackage{graphicx}

\begin{document}
\title{Band Alignment of 2D Semiconductors for Designing Heterostructures with Momentum Space Matching}

\author{V. Ongun {\"O}z{\c{c}}elik}\email{ongun@princeton.edu}
\affiliation{Andlinger Center for Energy and the Environment, Princeton University, Princeton, USA}
\author{Javad G. Azadani}
\affiliation{Department of Electrical and Computer Engineering, University of Minnesota, Minneapolis, USA}
\author{Ce Yang}
\affiliation{Institute of Microelectronics, Peking University, Beijing, China}
\author{Steven J. Koester}
\affiliation{Department of Electrical and Computer Engineering, University of Minnesota, Minneapolis, USA}
\author{Tony Low}\email{tlow@umn.edu}
\affiliation{Department of Electrical and Computer Engineering, University of Minnesota, Minneapolis, USA}

\begin{abstract}
We present a comprehensive study of the band alignments of two-dimensional (2D) semiconducting materials and highlight the possibilities of forming momentum-matched type I, II and III heterojunctions; an enticing possibility being atomic heterostructures where the constituents monolayers have band edges at the zone center, i.e. $\Gamma$ valley. Our study, which includes the Group IV and III-V compound monolayer materials, Group V elemental monolayer materials, transition metal dichalcogenides (TMD) and transition metal trichalcogenides (TMT) reveals that almost half of these materials have conduction and/or valence band edges residing at the zone center. Using first-principles density functional calculations, we present the type of the heterojunction for 903 different  possible combination of these 2D materials which establishes a periodic table of heterojunctions.
\end{abstract}

\maketitle

\section{Introduction}

Semiconductors have been at the heart of some of the most transformative device innovations over the course of the last 50 years, and research in two-dimensional (2D) atomic crystals has recently begun to focus on their heterostructures.\cite{geim2013van, novoselov2012two}  This includes heterostructures of graphene,\cite{novoselov2004electric, geim2007rise} which was followed by other monolayer structures each with different exceptional properties  such as the insulator boron-nitride;\cite{pacile2008two, dean2010boron} silicene and germanene which are the silicon and germanium based analogues of graphene;\cite{seymur2009two,lelay2012silicene,ozcelik2014new} oxygenated monolayers of graphene\cite{suk2010mechanical} and silicene,\cite{silicatene,yang2015ultrathin} transition-metal dichalcogenides (TMD),\cite{joensen1986single, ataca2011functionalization, radisavljevic2011single,tongay2012thermally} etc. Concomitantly, there are hundreds of different 2D materials and their permutations amount to numerous combinations of heterostructures. Certainly, theoretical exploration of these materials is needed to identify promising atomic heterostructures for device applications since depending on the field of usage, the requirements for heterostructures' band alignments change.

\begin{figure}
\includegraphics[width=8cm]{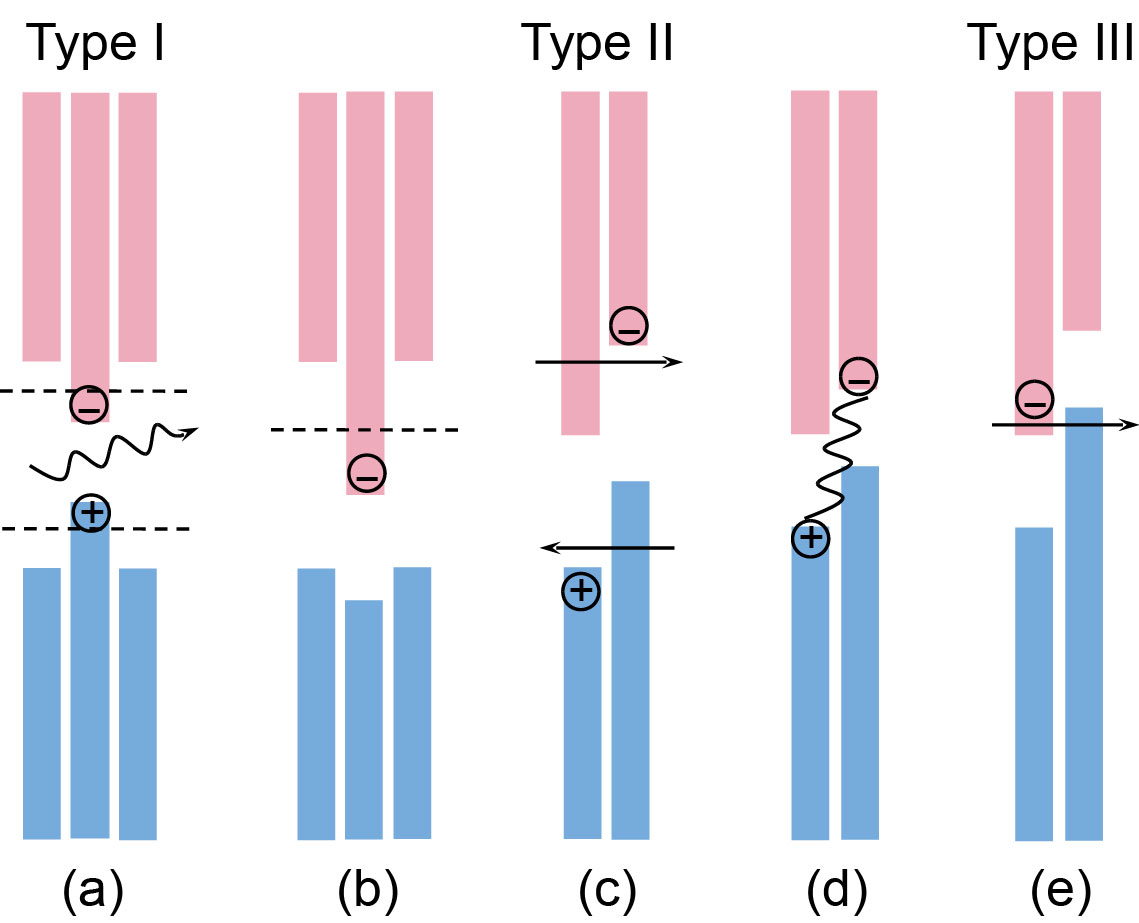}
\caption{Illustration of type I, II and III heterojunctions and their various devices, where red (blue) indicates conduction (valence) bands. (a) Laser utilizing type I heterojunction. (b) High electron mobility transistor, (c) carrier separation in solar cell and (d) interlayer excitons utilizing type II junction. (e) Tunneling field effect transistor based on type III junction.  In specific, when material A and material B merge, the resulting heterojunction is type I if VBM$_A$ $\textless$ VBM$_B$ $\textless$ CBM$_B$ $\textless$ CBM$_A$; is type II if VBM$_A$ $\textless$ VBM$_B$ $\textless$ CBM$_A$ $\textless$ CBM$_B$; and type III if VBM$_A$ $\textless$ CBM$_A$ $\textless$ VBM$_B$ $\textless$ CBM$_B$.}
\label{fig1}
\end{figure}

According to their bands alignments, heterojunctions can be classified into three types, i.e. type I (symmetric), type II (staggered), or type III (broken), as described in Fig.~\ref{fig1}.  Each of these band alignments has particular applications to enable different varieties of devices. Type I band alignments are most widely utilized in optical devices, such as light emitting diodes (LEDs)\cite{nakamura1995high} and in lasers as they provide a means to spatially confine electrons and holes so that efficient recombination can occur. The ability to fabricate single and multiple quantum wells has further enhanced the performance of lasers and light emitting diode devices.\cite{arakawa1986quantum, zory1993quantum} Type II band alignments are very useful for unipolar electronic device applications since they allow larger offsets on one side (either conduction or valence band), thus allowing extremely strong carrier confinement. Type II high electron mobility transistors (HEMTs) based upon InAs/AlSb quantum wells are excellent examples of the strong carrier confinement that can be achieved in type II heterostructures.\cite{werking1992high}. Conduction band notches can also be used as hot electron injector in bipolar transistors,\cite{levi1987room} and as quantum well resonant tunneling bipolar transistor.\cite{capasso1986quantum} Type II and type III heterojunctions are also useful to engineer the conduction to valence band transition energy.  This is particularly important in tunneling field effect transistors (TFETs) in order to enhance the tunneling current density,\cite{koswatta2010possibility} as well as in  infrared intersubband superlattice lasers,\cite{meyer1995type} and wavelength photodetectors.\cite{zhang2011long} Perhaps one of the most impressive devices enabled by semiconductor heterostructures are quantum cascade lasers, which use complex heterostructure stacks to engineer minibands and intrasubband transitions which allow efficient light emission from the mid-infrared to the terahertz regimes.\cite{faist1994quantum} Many optical device concepts have also been proposed or realized using 2D materials. Atomically-thin pn diodes of type II heterojunctions have recently been demonstrated to exhibit current rectification and collection of photoexcited carriers.\cite{lee2014atomically} The latter can also be achieved using graphene / TMD / graphene heterostructures.\cite{britnell2013strong} Phototransistor structures based on graphene / MoS$_2$ which exhibit ultra-high gain have also been realized.\cite{zhang2014ultrahigh} Light emitting diodes based on type I heterostructures with TMD sandwiched between BN and graphene as contacts can also be engineered across a wide spectral range.\cite{withers2015light} Very recently, type II heterostructures are also being explored as a platform for harnessing long-lived interlayer (or indirect) excitons.\cite{rivera2015observation, calman2015control} Additionally, lateral heterostructures of TMDs with other classes of monolayer materials can be utilized for various applications.\cite{ruzmetov2016vertical, yuan2015photoluminescence, gong2014vertical,ozcelik2015modulation}

Due to their extensive functionalities as described above, the electronic band structures of 2D materials and their bandgap engineering have been the focus of many recent studies. However, their relative band alignments (band offsets) have not been fully explored yet, except for the more common 2D materials such as graphene, BN and some TMDs.\cite{kim2015band, kang2013band, gong2013band} As far as heterostructures are concerned, the relative band alignment of semiconductors is one of the most important parameters of design since it identifies the type of the heterojunction. In addition, in all of the above-mentioned devices, just as in their 3D counterparts, it is critical for the conduction and/or valence band edges of constituent heterojunction materials to be momentum-matched. For instance, due to the inherent likelihood for some degree of mis-orientation in 2D layered stacks, and the desire to combine a highly diverse set of materials, 2D materials with band edges at the zone center  ($\Gamma$ point) point are extremely attractive, since momentum-space matching problems can largely be eliminated through the use of these materials.

\begin{figure*}
\includegraphics[width=14cm]{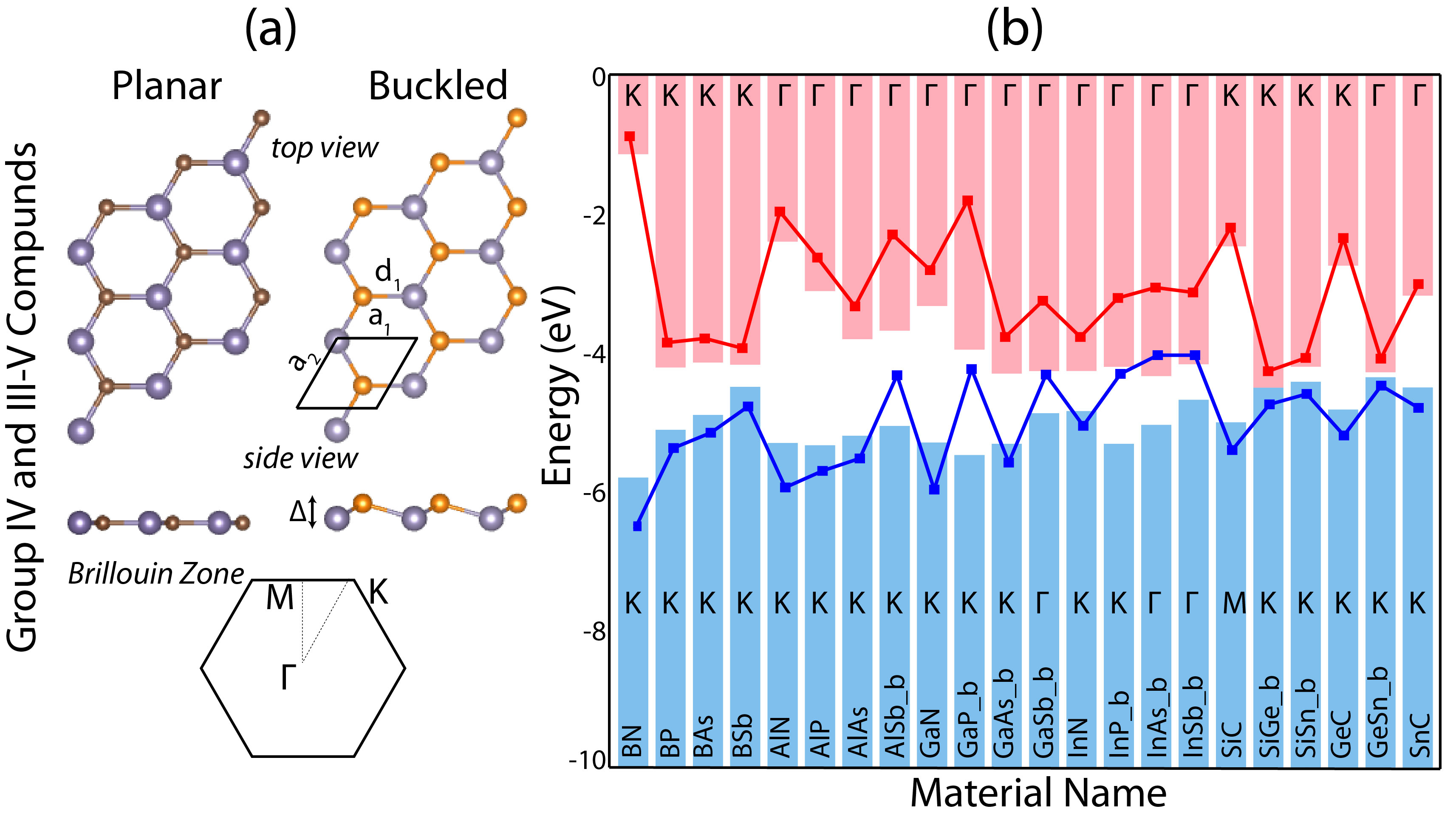}
\caption{(a) Stable crystalline structures (planar and buckled) of Group IV and Group III-V monolayer compounds, and their hexagonal BZ with high symmetry points are shown. The relevant bond lengths, lattice parameter and buckling distance are indicated by $a$, $d$, and $\Delta$, respectively. (b) Comparative band alignment of Group IV and Group III-V monolayer compounds where CBM (shown in red) and VBM (shown in blue) values obtained from PBE and HSE06 calculations are shown with bar and line plots, respectively.
The positions of CBM and VBM on the BZ are indicated.  Note that the vacuum energies were set to zero while calculating the band diagrams. Structures having buckled(b), puckered(w), 2H, 1T and distorted 1T phases are indicated with the corresponding indices, where no index is used for planar phases.}
\label{fig2}
\end{figure*}

In this letter, motivated by the emergence of 2D heterostructures as the next generation materials and the importance of comparative information of their band alignment and momentum matching for designing heterojunctions, we perform a comprehensive study of well-known 2D semiconducting materials, presenting their band edge properties and discuss the possibilities for forming momentum matched heterojunctions. Hence, our survey with density functional theory (DFT) calculations include various classes of 2D semiconductors such as Group IV and III-V compound monolayer materials, Group V elemental monolayer materials, TMDs and TMTs. For each class of semiconductors, we present the types of stable geometrical phases, structural dimensions, the electronic band structures, locations of the valance band maximum (VBM) and conduction band minimum (CBM) calculated with reference to vacuum in the momentum space, the values of the electronic affinities and the type of the band gap. These results establish a complete comparative database of 2D semiconductor materials whose band alignments and electronic properties are calculated using a unified approach. Following this, we calculate the type of the heterojunction for each possible combination of these 2D materials which results in a periodic table of heterojunctions. Thus, our letter presents a comprehensive library for determining the type of the heterojunction that will appear when two monolayer semiconductors are merged together.

\section{Method}

We performed first-principles pseudopotential calculations based on the spin-polarized DFT within generalized gradient approximation including van der Waals corrections \cite{grimme2006semi} and spin-orbit coupling. We used projector-augmented wave potentials\cite{blochl94} and approximated the exchange-correlation potential with Perdew-Burke-Ernzerhof (PBE) functional.\cite{pbe} We sampled the BZ in the Monkhorst-Pack scheme, and tested the convergence in energy as a function of the number of \textbf{k}-points for the calculations. The \textbf{k}-point sampling of (21$\times$21$\times$1) was found to be suitable for the Brillouin zone (BZ) corresponding to the primitive unit cell.  Atomic positions were optimized using the conjugate gradient method, where the total energy and atomic forces were minimized. The energy convergence value between two consecutive steps was chosen as $10^{-6}$ eV.  Numerical calculations were carried out using the VASP software\cite{vasp} where ``PREC=Accurate'' setting was used for structural minimization. Since the band gaps are usually underestimated by DFT, we also performed calculations using the HSE06 hybrid functional\cite{hse}, which is constructed by mixing 25\% of the Fock exchange with 75\% of the PBE exchange and 100\% of the PBE correlation energy. Electronic calculations at the HSE06 level were performed using the structures that were relaxed using PBE. Hence, our PBE and HSE06 results constitute our lower and upper bound estimates of the electronic gaps. The monolayer structures examined in this study were previously shown to be stable by means of phonon dispersion curves, high temperature molecular dynamics simulations or by experimental data; where relevant citations are given in the following sections for each group of materials. Throughout this work we present all of the energy values with reference to the vacuum energy, which we extracted from the local potential distribution within the unit cell. Thus, vacuum energies are set to zero in the band alignment figures. It should be noted that we present only the monolayer semiconductors where materials found to be metallic are not shown here.

\section{Band Alignments}
\subsection{Group IV and III-V compound materials}
After the exfoliation of hexagonal boron nitride monolayers,\cite{novoselov2005two, pacile2008two} search for similar Group IV and III-V compound materials and designing nanoscale devices composed of their heterostructures has drawn considerable attention.\cite{emtsev2009towards,ozcelik2013nanoscale,kecik2015layer,ozccelik2015high,onen2016gan}  It is possible to separate these materials into two distinct groups depending on their stable geometries. \cite{csahin2009monolayer} The compounds in the first group (namely BN, BP, BAs, BSb,AlN, AlP, AlAs, GaN, InN, SiC, GeC, SnC) have the same hexagonal honeycomb structure similar to graphene but with different lattice constants. Among these, BN is most closely lattice matched with graphene having a lattice constant of 1.45 \AA~ (versus 1.42 \AA~ for graphene), but with ionic bonds and a wide band gap as opposed to the zero band gap graphene. The other stable structure that Group IV and III-V compounds possess is the buckled geometry where the adjacent atoms of the material are in two parallel planes separated by a buckling distance ($\Delta$) in the vertical direction, as shown in Fig.~\ref{fig2}(a). It should be noted that compounds with a  buckled geometry (namely: AlSb, GaP, GaAs, GaSb, InP, InAs, InSb, SiGe, SiSn, GeSn) do not contain an element from the second row of the periodic table (as opposed to planar structures) and their stability is maintained by the buckling of the bonds. Despite this buckling in the vertical direction, these materials maintain their hexagonal symmetries and all have the same hexagonal BZ as their planar counterparts.

Monolayers of Group IV and III-V compounds include a variety of different band alignments ranging from direct to indirect gap semiconductors with VBM and CBM at K, $\Gamma$, and M points in the momentum space, as illustrated in Fig.~\ref{fig2}(b). There are also wide band-gap insulators among these materials such as BN, AlN and SiC.  As a general trend, the VBM (CBM) increases (decreases) as the row numbers of the elements in the compounds increase; resulting in a corresponding decrease in the band gap. Among these, GaN and AlN which have similar lattice constants, are attractive candidates for both lateral and vertical heterostructures. Additionally, buckled compounds of AlSb, GaP, GaSb, InP, InAs and InSb have very high VBM, approaching -4eV. These can be combined with low CBM materials to form type III heterojunctions. Also, large gap compounds, like BN, are natural candidates to combine with smaller gap compound (like SiGe) to achieve a type I heterojunctions. Specific candidates can be identified using the complete set of data presented in Table 1.

\subsection{Group V monolayers}
\begin{figure}
\includegraphics[width=8cm]{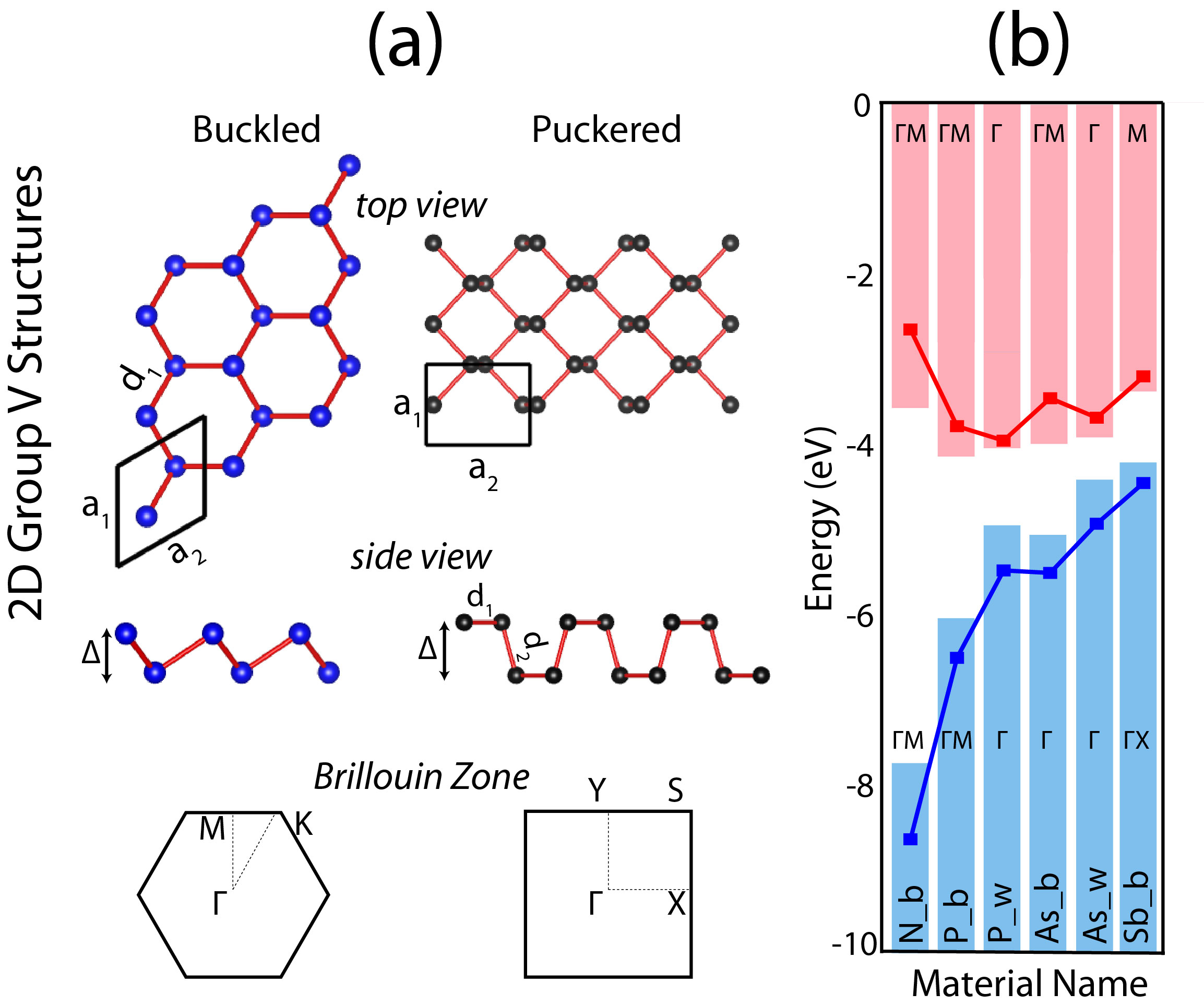}
\caption{(a) Stable crystalline structures (buckled and puckered) of Group V elements, and their corresponding hexagonal and rectangular BZ with high symmetry points are shown. The relevant bond lengths and lattice parameters are indicated by $a$ and $d$, respectively. (b) Same as Figure 2(b), but for Group V monolayers.}
\label{fig3}
\end{figure}

Recently, another class of 2D materials, the Group V monolayers, has also garnered attention in the community. Specially, the demonstration of field effect transistors using multilayers of black phosphorene \cite{li2014black, liu2014phosphorene} and related theoretical studies \cite{zhu2014semiconducting,low2014tunable,rodin2014strain} have brought Group V elements into focus. More recently, theoretical and experimental studies have been performed to explore the possible stable phases of other Group V elements such as nitrogen, antimony, arsenic and bismuth.\cite{ozccelik2015prediction, akturk2015single, zhang2015atomically, kamal2015arsenene, freitas2015topological} Reported high temperature molecular dynamics simulations and DFT calculations in these previous studies show the existence of two distinct geometries. As shown in Fig.~\ref{fig3}(a), the first geometry is the buckled structure similar to that observed in some of Group IV and III-V compounds. All of the Group V elements have a stable monolayer phase in this buckled geometry. Among these, monolayer nitrogen (or nitrogene) is a wide band gap insulator with a fundamental direct band gap of 4.0 eV between $\Gamma$ and M symmetry points. However, this band gap increases to 5.9 eV when calculated by HSE06 funcional.\cite{hse} Similarly, the buckled structure of phosphorous, namely blue phosphorene,\cite{liu2014phosphorene,zhu2014semiconducting} also has a direct band gap between $\Gamma$ and M symmetry points, with a value of 1.98 eV which increases to 2.73 eV upon HSE06 corrections. In contrast to buckled nitrogene and phosphorene, the buckled monolayers of arsenic and antimony (arsenene and antimonene) have indirect band gaps. Apart from these buckled structures; phosphorene, arsenene and antimonene have a different form of stable monolayer, namely the washboard (or puckered) geometry, which is also illustrated in Fig.~\ref{fig3}(a). This is the phosphorene phase that most experiments reported recently as black phosphorene. \cite{kou2015phosphorene, liu2014phosphorene,wu2015atomic,li2014black} In the puckered phase, the structures lose their hexagonal symmetries and both their unit cell and BZ become rectangular. When stabilized in this geometry, the band gaps of phosphorene and arsenene decrease significantly as the VBM values increase and the band gap location of puckered phosphorene shifts to the $\Gamma$ point. Finally, it should be noted that in addition to their monolayers, these Group V materials can also form stable bilayer and layered 3D structures, where the puckered antimonene has a slightly distorted phase\cite{ozccelik2015prediction, akturk2015single}

In Fig.~\ref{fig3}(b) we present the complete band alignment properties of Group V monolayers. In particular, puckered phosphorene and arsenene retain their $\Gamma$ valley band edges even with increasing layer numbers. One can envision various type I or II $\Gamma$ valley lateral heterostructures by controlling layer numbers.\cite{luo2016thermal} In addition, these Group V puckered elements tend to have relatively low CBMs, $\sim$ -4eV which allows them to form type I or II heterostructures with  Group IV and III-V compounds that we discussed earlier; particularly those with high VBM of $\sim$ -4eV.

\subsection{Transition metal dichalcogenides}
\begin{figure}
\includegraphics[width=8cm]{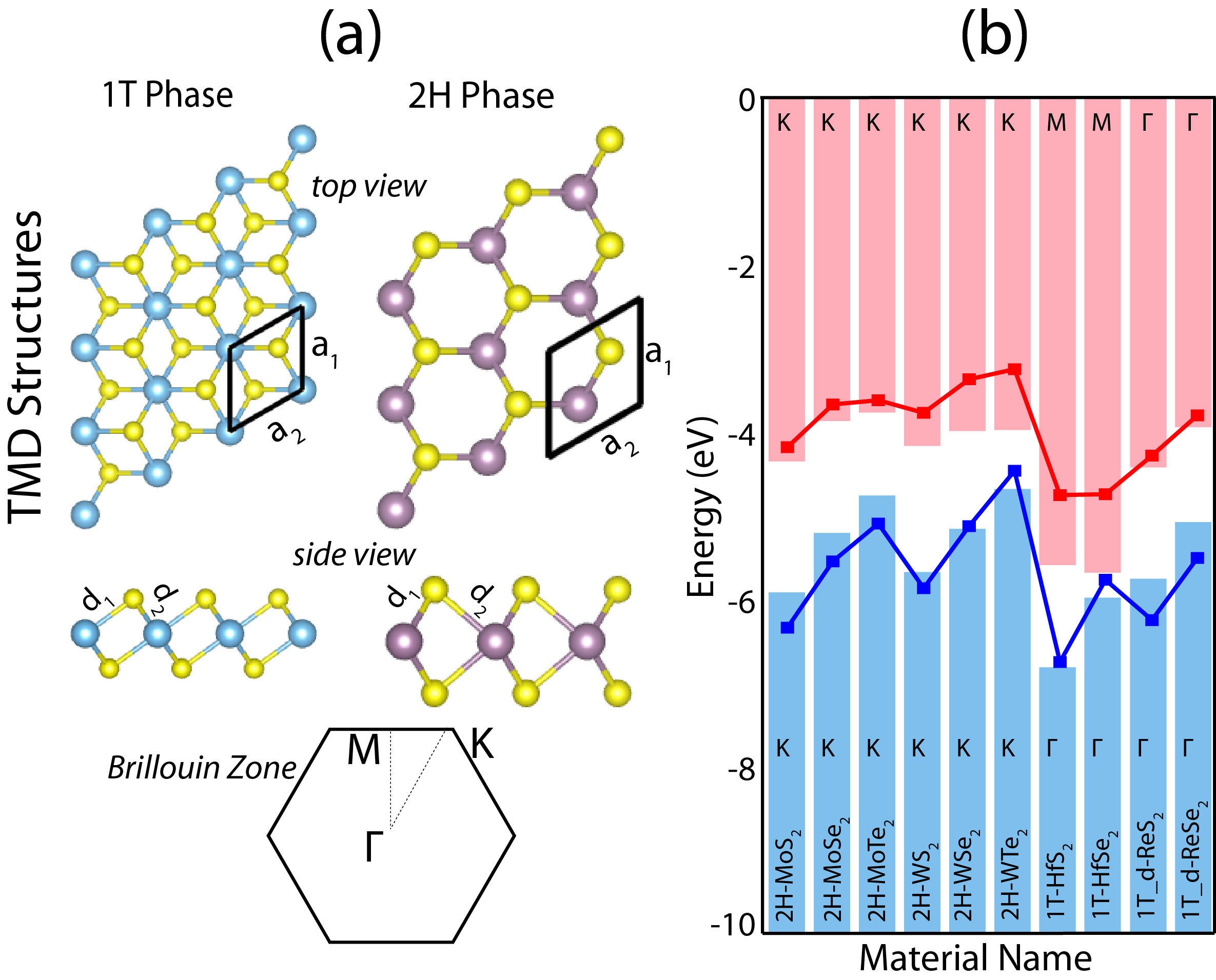}
\caption{(a) Stable crystalline structures (1T and 2H phases) of TMDs, and their corresponding hexagonal BZ with high symmetry points are shown. The relevant bond lengths and lattice parameters are indicated by $a$ and $d$, respectively. Note that the structures of ReS$_2$ and ReSe$_2$ are in a distorted 1T phase. (b) Same as Figure 2(b) but for TMDs.}
\label{fig4}
\end{figure}

So far we have discussed the monolayer structures of Group III, IV and V elements and their compounds. Additionally, it is also possible to construct stable 2D materials using transition metals to form TMDs, which exhibit a versatile chemistry.\cite{joensen1986single, coleman2011two, ataca2011functionalization} The general formula of a TMD is MX$_2$, in which M stands for a transition metal atom and X is a chalcogen (S, Se, Te).\cite{chhowalla2013chemistry} Most common TMDs are either in 1T or 2H phases, where 1T and 2H refer to the structure of the lattices as exhibited in  Fig.~\ref{fig4}(a). In the 1T-MX$_2$ phase, the lattice is octahedral (Oh), while in the 2H-MX$_2$ phase it is trigonal prismatic (D3h). In addition, ReS$_2$ and ReSe$_2$ are observed with a distorted 1T crystal structure due to Peierls or Jahn-Teller distortions.\cite{tongay2014monolayer, wolverson2014raman} As shown in Fig.~\ref{fig4}(b), monolayer MX$_2$ structures (M = Mo, W) are typically direct-band-gap semiconductors, whereas  TiX$_2$ is metallic, hence not shown in this work. With the increase in the number of layers, the electronic structure of 2H-MX$_2$ changes from direct to indirect band gap and the band gap also decreases.\cite{yun2012thickness, wang2012electronics}

The relaxed geometries of TMDs show that as X goes from S to Te, the bond lengths and lattice constants increase slightly. Also, among these materials, MoX$_2$ and WX$_2$ have the closest lattice constants to each other, suggesting possible applications in various lateral heterostructures. In Fig.~\ref{fig4}(b), we present the band alignment results of all of the stable MX$_2$ semiconductor phases. As a general trend, the VBM and CBM  increase as X changes from S to Te, similar to the trend in the lattice constants. Also, it should be noted that among structures containing the same X atom, the CBM and VBM value of WX$_2 $ is always the highest. This suggests the construction of type II heterojunctions; for example by using WX$_2$ and MoX$_2$. As far as the k-space location of the band gaps are concerned, all of the M=Mo and M=W structures have direct band gaps at the K point. For the remaining materials, the VBM is always at the $\Gamma$ point and CBM is at the M point if M=Hf; and at $\Gamma$ for M=Re. It should be noted that for M=Re we have semiconductors with direct band gap at the $\Gamma$ point. These Re based TMDs were also found to have a band gap that is relatively insensitive to the number of layers.\cite{tongay2014monolayer}

\subsection{Transition metal trichalcogenides}
\begin{figure}
\includegraphics[width=8cm]{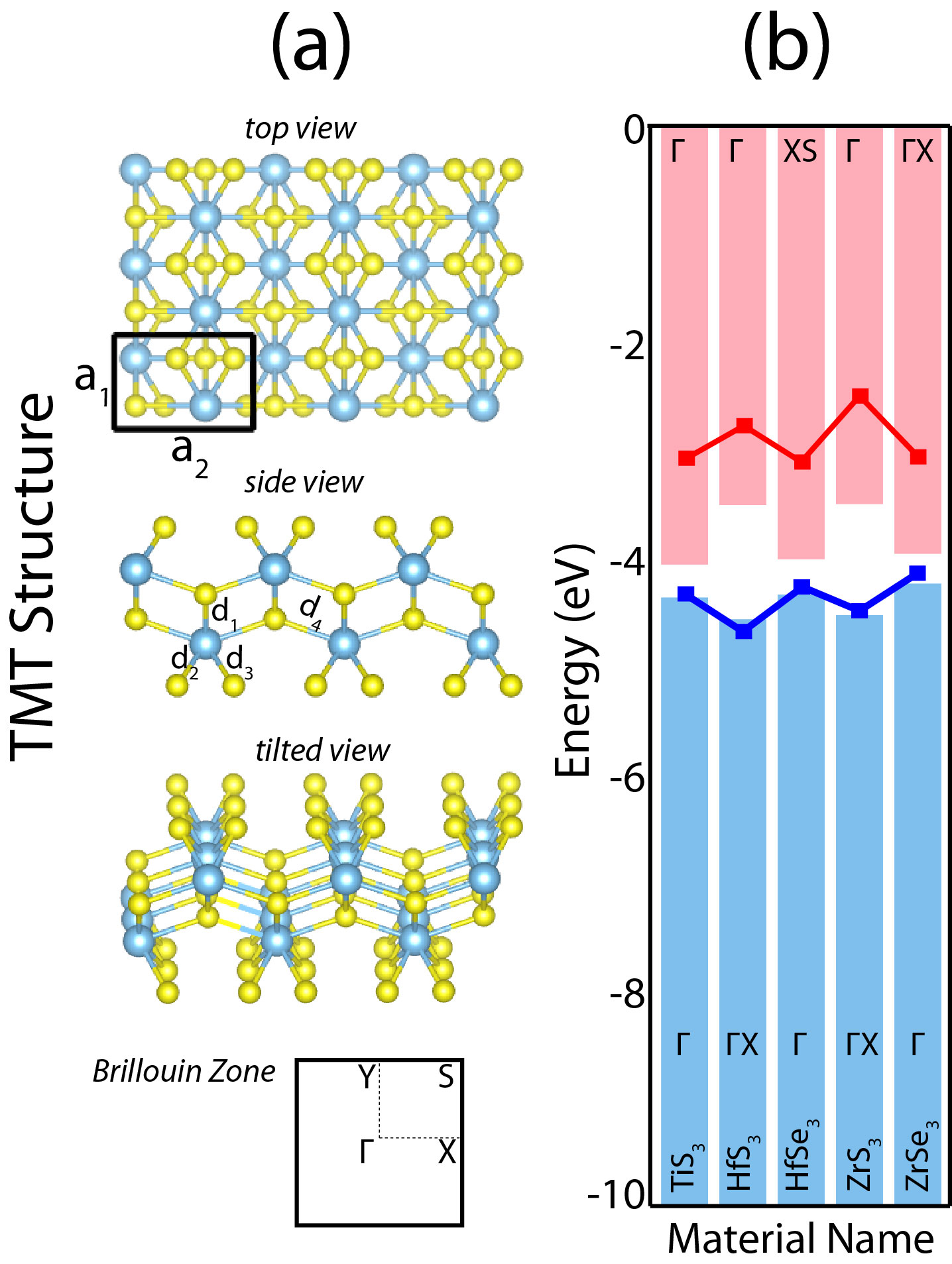}
\caption{ (a) Stable crystalline structure of TMTs, and its corresponding rectangular BZ with high symmetry points are shown. The relevant bond lengths and lattice parameters are indicated by $a$ and $d$, respectively. (b) Same as Figure 2(b) but for TMTs.}
\label{fig5}
\end{figure}

Another family of semiconducting monolayer materials is the transition metal trichalcogenides (TMT) which are layered structures with weak interlayer van der Waals interactions. The general chemical formula for a TMT is MX$_3$, where M is the transition metal Ti, Zr or Hf and X is a chalcogen, i.e. S, Se or Te.  They have monoclinic crystalline structures\cite{furuseth1975crystal, brattas1972properties} with rectangular unit cells containing two metal and six chalcogen atoms, where each metal atom is  connected to six chalcogen atoms as shown in Fig.~\ref{fig5}(a). Previous experimental studies have reported the electronic properties of bulk TMTs.\cite{finkman1984electrical, kikkawa1980electrical, gorlova2010features, srivastava1992preparation, levy1983single}

\begin{figure*}
\includegraphics[width=14cm]{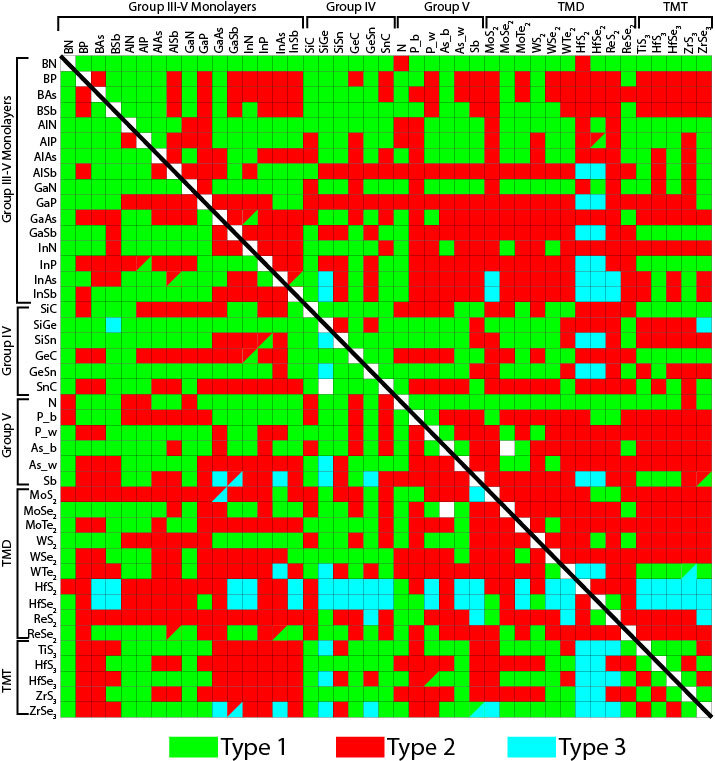}
\caption{ Periodic table of heterojunctions. Type I, II and III heterojunctions are represented by green, red and blue boxes, respectively. The lower left and upper right of regions of the diagonal line present results calculated by PBE and HSE06, respectively. Two-colored boxes indicate equal chances of two different types of heterojunctions for the corresponding materials.}
\label{fig6}
\end{figure*}

In Fig.~\ref{fig5}(b), we present the band alignment results for this family of monolayer semiconductors. Our calculations reveal that, among the semiconducting monolayers of TMTs, only TiS$_3$ is a direct band gap semiconductor where both the VBM and CBM are located at the $\Gamma$ point. Using PBE functional,\cite{pbe} TiS$_3$ has a fundamental gap of 0.32 eV and this increases to 1.04 eV when calculated by HSE06, which is comparable with the experimentally observed value of 1.10 eV. \cite{ferrer2013optical, ferrer2012photoelectrochemical} Similarly, the band gaps of all TMTs studied here increase and get closer to their experimentally observed values when calculated with HSE06. They are all indirect gap semiconductors, having either the VBM or the CBM slightly offset from the $\Gamma$ point. TMTs have low CBM of about -4.5eV, so they can be combined with high VBM monolayers, such as HfS$_2$ and HSe$_2$ to form type III heterojunctions.  It is also possible to construct type II heterostructures by combining appropriate TMTs with MoS$_2$ and ReS$_2$. Combining HfS$_3$ with HfSe$_2$ would make a type II heterojunction. Here we note that, for HfSe$_3$ the location of the CBM changes from XS to X when calculated by HSE06.

\section{Periodic Table of Heterojunctions}

Having calculated the geometrical and electronic properties of all well-known two dimensional semiconductors using a unified approach, we finally focus on establishing the complete set of heterojunction types that would form when any two of these materials are merged together; some of which were discussed above. For this, we use the criteria illustrated in Fig.~\ref{fig1}. Namely, when material A and material B merges, the resulting heterojunction is type I if VBM$_A$ $\textless$ VBM$_B$ $\textless$ CBM$_B$ $\textless$ CBM$_A$; is type II if VBM$_A$ $\textless$ VBM$_B$ $\textless$ CBM$_A$ $\textless$ CBM$_B$; and type III if VBM$_A$ $\textless$ CBM$_A$ $\textless$ VBM$_B$ $\textless$ CBM$_B$. Using these criteria, we systematically present our results in the periodic table of heterojunctions in Fig.~\ref{fig6}, using both the PBE and HSE06 approaches. Accordingly, we mostly see type I heterojunctions in Group III-V and Group IV compounds, where as type II heterojunctions dominate as we move to Group V, TMD and TMT regions. Also, type III heterojunctions emerges in TMD and TMT regions of the periodic table. It should be noted that, for some rare cases the VBM of material A is equal to the CBM of material B (or vice versa); which suggests that the type of the heterojunction can vary between two possibilities depending on other conditions, which are beyond the scope of this study. We also note that the presented results also depend on the type of density functional (i.e. PBE versus HSE06) used. In addition, the band alignments might also be modified in the presence of interface dipoles in the heterostructures, an effect we will defer to a future study. Nevertheless, the results presented here in the periodic table of heterojunctions provide a useful guidance for experimentalists in selecting suitable momentum-matched 2D materials for heterojunction design.

\section{Conclusion}
In conclusion, we have presented a complete and comparative set of properties of well-known semiconductors including their structural and electronic properties. We identified possible types of heterojunctions that can be realized using these materials. Using these unified results, we constructed a periodic table of heterojunctions which provides a useful guidance for future experimental / theoretical studies.  Additionally, the presented band alignment results and their band edges in momentum space using a consistent methodology provide a rich database for creating heterojunctions with momentum space matching. In particular, we found that about half of the 2D semiconductors we surveyed have either (or both) the conduction and valence band edges at the zone center $\Gamma$, hence they can be excellent candidates in constructing momentum-matched heterostructures. We believe that the suggested examples in this letter, along with many others, can be constructed using our unified periodic table of heterojunctions.

\section{Acknowledgements}
We acknowledge fruitful discussion with Dr. Mehmet Topsakal. JGA and TL acknowledge partial support from NSF ECCS-1542202. SJK was supported in part by the National Science Foundation (NSF) through the University of Minnesota MRSEC under Award DMR-1420013. The computational resources were provided by the Minnesota Supercomputing Institute (MSI).

\section{Supporting Information}
The electronic band diagrams of the materials studied in this paper are available in the supporting information.

\begin{table*}
\small
\caption{Relevant lattice constants, bond lengths, buckling distance, VBM and CBM values for heterostructures studied are shown. Structures having buckled(b), puckered(w), 2H, 1T and distorted 1T phases are indicated with the corresponding indices, where no index is used for planar phases. The listed geometrical parameters are illustrated in Figures 2, 3, 4 and 5. The VBM and CBM values calculated with HSE are also shown in parentheses. }
\begin{tabular}{cccccc}
\hline
Material &  Lattice Constants (\AA) & Bond Lengths (\AA) & Buckling (\AA) & VBM (eV) & CBM(eV)  \tabularnewline
\hline
 BN &  a$_1$=a$_2$=2.51 & d$_1$=1.45 & $\Delta$=0 & -5.80 (-6.56) & -1.14 (-0.88) \tabularnewline
 BP &  a$_1$=a$_2$=3.21 & d$_1$=1.85 & $\Delta$=0 & -5.08 (-5.43) & -4.18 (-3.84) \tabularnewline
 BAs &  a$_1$=a$_2$=3.39 & d$_1$=1.96 & $\Delta$=0 & -4.89 (-5.21) & -4.13 (-3.78) \tabularnewline
 BSb &  a$_1$=a$_2$=3.73 & d$_1$=2.15 & $\Delta$=0 & -4.49 (-4.83) & -4.17 (-3.92) \tabularnewline
 AlN &  a$_1$=a$_2$=3.12 & d$_1$=1.80 & $\Delta$=0 & -5.29 (-6.00) & -2.38 (-1.96) \tabularnewline
 AlP &  a$_1$=a$_2$=3.95 & d$_1$=2.28 & $\Delta$=0 & -5.32 (-5.76) & -3.09 (-2.62) \tabularnewline
 AlAs &  a$_1$=a$_2$=4.07 & d$_1$=2.35 & $\Delta$=0 & -5.16 (-5.58) & -3.76 (-3.23) \tabularnewline
 AlSb\textunderscore b &  a$_1$=a$_2$=4.41 & d$_1$=2.62 & $\Delta$=0.62 & -5.02 (-4.38) & -3.64 (-2.29) \tabularnewline
 GaN &  a$_1$=a$_2$=3.25 & d$_1$=1.88 & $\Delta$=0 & -5.24 (-6.03) & -3.27 (-2.80) \tabularnewline
 GaP\textunderscore b &  a$_1$=a$_2$=3.89 & d$_1$=2.30 & $\Delta$=0.48 & -5.46 (-4.29) & -3.93 (-1.80) \tabularnewline
 GaAs\textunderscore b &  a$_1$=a$_2$=4.04& d$_1$=2.41 & $\Delta$=0.61 & -5.27 (-5.64) & -4.25 (-3.76) \tabularnewline
 GaSb\textunderscore b &  a$_1$=a$_2$=4.43 & d$_1$=2.64 & $\Delta$=0.68 & -4.85 (-4.37) & -4.24 (-3.24) \tabularnewline
 InN &  a$_1$=a$_2$=3.58 & d$_1$=2.06 & $\Delta$=0 & -4.81 (-5.11) & -4.23 (-3.76) \tabularnewline
 InP\textunderscore b &  a$_1$=a$_2$=4.27 & d$_1$=2.53 & $\Delta$=0.56 & -5.32 (-4.36) & -4.21 (-3.20) \tabularnewline
 InAs\textunderscore b &  a$_1$=a$_2$=4.35 & d$_1$=2.61 & $\Delta$=0.71 & -5.02 (-4.09) & -4.31 (-3.05) \tabularnewline
 InSb\textunderscore b &  a$_1$=a$_2$=4.66 & d$_1$=2.80 & $\Delta$=0.77 & -4.68 (-4.09) & -4.16 (-3.12) \tabularnewline
 SiC &  a$_1$=a$_2$=3.09 & d$_1$=1.78 & $\Delta$=0 & -5.01 (-5.46) & -2.46 (-2.19) \tabularnewline
 SiGe\textunderscore b &  a$_1$=a$_2$=3.94 & d$_1$=2.35 & $\Delta$=0.60 & -4.52 (-4.80) & -4.51 (-4.25) \tabularnewline
 SiSn\textunderscore b &  a$_1$=a$_2$=4.26 & d$_1$=2.55 & $\Delta$=0.69 & -4.43 (-4.65) & -4.21 (-4.06) \tabularnewline
 GeC &  a$_1$=a$_2$=3.25 & d$_1$=1.88 & $\Delta$=0 & -4.81 (-5.25) & -2.73 (-2.34) \tabularnewline
 GeSn\textunderscore b &  a$_1$=a$_2$=4.42 & d$_1$=2.66 & $\Delta$=0.75 & -4.35 (-4.53) & -4.27 (-4.07) \tabularnewline
 SnC &  a$_1$=a$_2$=3.58 & d$_1$=2.07 & $\Delta$=0 & -4.51 (-4.85) & -3.53 (-3.00) \tabularnewline
 N\textunderscore b &  a$_1$=a$_2$=2.27 & d$_1$=1.49 & $\Delta$=0.70 & -7.66 (-8.68) & -3.57 (-2.67) \tabularnewline
 P\textunderscore b &  a$_1$=a$_2$=3.28 & d$_1$=2.26 & $\Delta$=1.24 & -6.09 (-6.54) & -4.11 (-3.81) \tabularnewline
 P\textunderscore w &  a$_1$=4.55; a$_2$=3.31 & d$_1$=2.22; d$_2$=3.52 & $\Delta$=2.11 & -4.87 (-5.51) & -4.01 (-3.98) \tabularnewline
 As\textunderscore b &  a$_1$=a$_2$=3.61 & d$_1$=2.51 & $\Delta$=1.40 & -5.13 (-5.54) & -3.95 (-3.48) \tabularnewline
 As\textunderscore w &  a$_1$=4.72; a$_2$=3.67 & d$_1$=2.51; d$_2$=3.84 & $\Delta$=2.41 & -4.41 (-4.96) & -3.86 (-3.71) \tabularnewline
 Sb\textunderscore b &  a$_1$=a$_2$=4.04 & d$_1$=2.87 & $\Delta$=1.68 & -4.24 (-4.48) & -3.46 (-3.26) \tabularnewline
 2H-MoS$_2$ &  a$_1$=a$_2$=3.19 & d$_1$=2.41; d$_2$=4.00 & - & -5.84 (-6.33) & -4.25 (-4.18) \tabularnewline
 2H-MoSe$_2$ &  a$_1$=a$_2$=3.32 & d$_1$=2.54; d$_2$=4.18 & - & -5.13 (-5.54) & -3.79 (-3.67) \tabularnewline
 2H-MoTe$_2$ &  a$_1$=a$_2$=3.55 & d$_1$=2.73; d$_2$=4.48 & - & -4.64 (-5.09) & -3.68 (-3.62) \tabularnewline
 2H-WS$_2$ &  a$_1$=a$_2$=3.19 & d$_1$=2.42; d$_2$=4.00 & - & -5.36 (-5.86) & -3.82 (-3.77) \tabularnewline
 2H-WSe$_2$ &  a$_1$=a$_2$=3.32 & d$_1$=2.55; d$_2$=4.19 & - & -4.70 (-5.12) & -3.47 (-3.37) \tabularnewline
 2H-WTe$_2$ &  a$_1$=a$_2$=3.56 & d$_1$=2.74; d$_2$=4.49 & - & -4.29 (-4.46) & -3.55 (-3.25) \tabularnewline
 1T-HfS$_2$ &  a$_1$=a$_2$=3.64 & d$_1$=2.55; d$_2$=4.45 & - & -6.18 (-6.74) & -4.95 (-4.75) \tabularnewline
 1T-HfSe$_2$ &  a$_1$=a$_2$=3.76 & d$_1$=2.68; d$_2$=4.62 & - & -5.33 (-5.76) & -4.93 (-4.74) \tabularnewline
 1T\textunderscore d-ReS$_2$ &  a$_1$=6.38; a$_2$=6.47 & d$_1$=2.51; d$_2$=2.38 & - & -5.73 (-6.24) & -4.40 (-4.28) \tabularnewline
 1T\textunderscore d-ReSe$_2$ &  a$_1$=6.58; a$_2$=6.74 & d$_1$=2.51; d$_2$=2.51 & - & -5.02 (-5.50) & -3.90 (-3.80) \tabularnewline
 TiS$_3$ &  a$_1$=4.99; a$_2$=3.39 & d$_1$=2.45; d$_2$=d$_3$=2.49; d$_4$=2.65 & - & -4.37 (-4.33) & -4.05 (-3.29) \tabularnewline
 HfS$_3$ &  a$_1$=5.09; a$_2$=3.58 & d$_1$=2.55; d$_2$=d$_3$=2.60; d$_4$=2.68 & - & -4.57 (-4.59) & -3.51 (-2.72) \tabularnewline
 HfSe$_3$ &  a$_1$=5.40; a$_2$=3.71 & d$_1$=2.69; d$_2$=d$_3$=2.74; d$_4$=2.85 & - & -4.34 (-4.30) & -4.01 (-3.32) \tabularnewline
 ZrS$_3$ &  a$_1$=5.14; a$_2$=3.62 & d$_1$=2.60; d$_2$=d$_3$=2.62; d$_4$=2.72 & - & -4.53 (-4.46) & -3.50 (-2.57) \tabularnewline
 ZrSe$_3$ &  a$_1$=5.42; a$_2$=3.74 & d$_1$=2.74; d$_2$=d$_3$=2.76;  d$_4$=2.88 & - & -4.24 (-4.21) & -3.96 (-3.26) \tabularnewline
\hline
\end{tabular}
\end{table*}

\bibliography{arxiv.bbl}

\end{document}